\documentclass[aps,prl,twocolumn,notitlepage,groupedaddress]{revtex4-1}

\usepackage[utf8]{inputenc}
\usepackage{amsmath,amsfonts,amssymb}
\usepackage{graphicx}

\usepackage{hyperref}
\usepackage[usenames,dvipsnames]{xcolor}

\usepackage{tikz}
\hypersetup{colorlinks=true, linkcolor=BrickRed, urlcolor=blue!50!black, citecolor=blue!50!black}

\newcommand\diff{\mathrm{d}}

\renewcommand{\vec}[1]{\mathbf{#1}}
\renewcommand{\phi}[0]{\varphi}

\usepackage[normalem]{ulem}

\begin{document}

\title{Diverging time scale in the dimensional crossover for liquids in strong confinement}
\author{Suvendu Mandal}
\affiliation{Institut f\"ur Theoretische Physik, Universit\"at Innsbruck, Technikerstr. 21A, A-6020 Innsbruck, Austria}
\author{Thomas Franosch}
\affiliation{Institut f\"ur Theoretische Physik, Universit\"at Innsbruck, Technikerstr. 21A, A-6020 Innsbruck, Austria}

\begin{abstract}

We study a strongly interacting dense hard-sphere system confined between two parallel plates by event-driven molecular dynamics simulations
to address the fundamental question of the nature of the 3D to 2D crossover. As the fluid becomes more and more confined 
the dynamics of the transverse and lateral degrees of freedom decouple, which is accompanied by a diverging time scale 
separating 2D from 3D behavior. Relying on the time-correlation function of the  transversal kinetic energy the scaling behavior  
and its density-dependence is explored. Surprisingly, our simulations reveal that its time-dependence becomes purely exponential 
such that memory effects can be ignored. We rationalize our findings quantitatively in terms of an analytic theory 
which becomes exact in the limit of strong confinement.

\end{abstract}

\date{\today}
\maketitle

\paragraph{Introduction.---}

Transport of particles in nanoconfinement is of great scientific and industrial importance with applications in 
heterogeneous catalysis~\cite{Sheehan:2013}, oil recovery~\cite{Sahimi:1993}, 
or  lubrication~\cite{Granick:1991,Fluid_film_lubrication:2010,Santana-Solano:2005,Ekiel:2008}. In recent years artificial 
nanoporous materials such as metal organic 
frameworks~\cite{Rosi:2003,Quah:2015}, zeolites~\cite{Tomita:2004,Karger_Diffusion}, and biocompatible 
scaffolds~\cite{Wang:2013} have also triggered many 
novel applications, including gas storage~\cite{Schlapbach:2001}, repairing or regenerating tissues~\cite{Wang:2013}, size-selective molecular sieving~\cite{Han:2008}, lab-on-a chip technology 
and nanofluidics~\cite{Eijkel:2005,Mijatovic:2005}. The efficiency of such nanodevices often crucially depends on higher surface to volume 
ratio, such that the distance between the confining walls may even reach atomic scale~\cite{Lien:2015}, or the system effectively becomes 
quasi-2D. Nevertheless, despite its long history, a deep understanding of the transport mechanisms in nanoconfinement 
and how the dimensional crossover occurs dynamically is 
still far from satisfactory. 

Early theoretical studies on transport in nanoconfinement starting from Knudsen and Smoluchowski~\cite{Knudsen:1909,Smoluchowski:1910} focused  
on dilute  hard-sphere
gases where exact results could be obtained analytically in the 
low-density limit~\cite{Knudsen:1909,Smoluchowski:1910,Arya:2003,Gruener:2008,Malek:2001,Kim:2008} by assuming particle-wall 
collisions as diffusive.
In contrast, confinement effects on dense  strongly interacting systems have only recently come into focus~\cite{Lowen:2009}. There, the simplest geometry to investigate the effects of strong confinement is a slit where fluid particles are restricted to a narrow space between two smooth parallel plates, but also tubes or spherical confinements have been realized experimentally~\cite{Saklayen:2013,Hunter:2014,Bo:2016}.
 Computer simulations and experiments for the planar confinement have revealed  an exotic equilibrium phase behavior due to commensurable stacking~\cite{Schmidt:1996,*Schmidt:1997,Gribova:2011,Fortini:2006,Fontecha:2005, Oguz:2009, *Oguz:2012a, *Oguz:2012b, Reinmueller:2013b,Varnik:2016,Alba:2006} as well as the hexatic phases in the limit of quasi-2D confinement~\cite{Qi:2015,Li:2016}. 
Confinement induced order-disorder phase transitions for certain nonpolar liquids  have also been reported in several experiments~\cite{Klein:1995,*Klein:1998a,*Klein:1998b}, but the interpretation has been challenged in favor of a glass transition~\cite{Kienle:2016,Demirel:2001,Zhu:2004}. The structural properties of strongly confined liquids have been measured directly only recently by x-ray scattering~\cite{Nugent:2007,* Nygard:2012, *Nygard:2016a,*Nygard:2016b, Satapathy:2008, *Satapathy:2009}. 

The structural changes by the confinement also have drastic ramifications for the dynamic properties. For instance, the role of local order has been elucidated within a remarkable empirical scaling of the diffusivity or structural relaxation times with the excess entropy~\cite{Mittal:2006,*Mittal:2007a,*Mittal:2007b,*Mittal:2008,Ingebrigtsen:2013, Goel:2008, *Goel:2009, *Krekelberg:2013,Fehr:1995}. Complementarily, 
a microscopic theory for the dynamics in confinement is the mode-coupling theory that predicts a multiple-reentrant in a glassy phase as the plate separation is varied~\cite{Lang:2010, *Lang:2012, *Lang:2013, *Lang:2014a, Mandal:2014, Varnik:2016}

From a more fundamental point of view one would like to know how the dimensional crossover from a 3D bulk liquid to a (quasi-)2D system occurs. 
For the thermodynamic and structural properties it has been shown recently that a small parameter emerges such that the convergence to a 2D system including leading corrections  can be proven~\cite{Franosch:2012,*Lang:2014}. The key observation there was that in strong confinement 
the canonical ensemble for the fluid in a slit geometry decouples into a two-dimensional fluid in the lateral plane and an ideal gas in the transversal direction. However, the consequences of 
the weak coupling between lateral and transversal degrees of freedom for  slow equilibration 
and how time-dependent correlation functions will be affected by coupling to the 'other dimension' have remained mostly unexplored.


\begin{figure}
\includegraphics*[width=\linewidth]{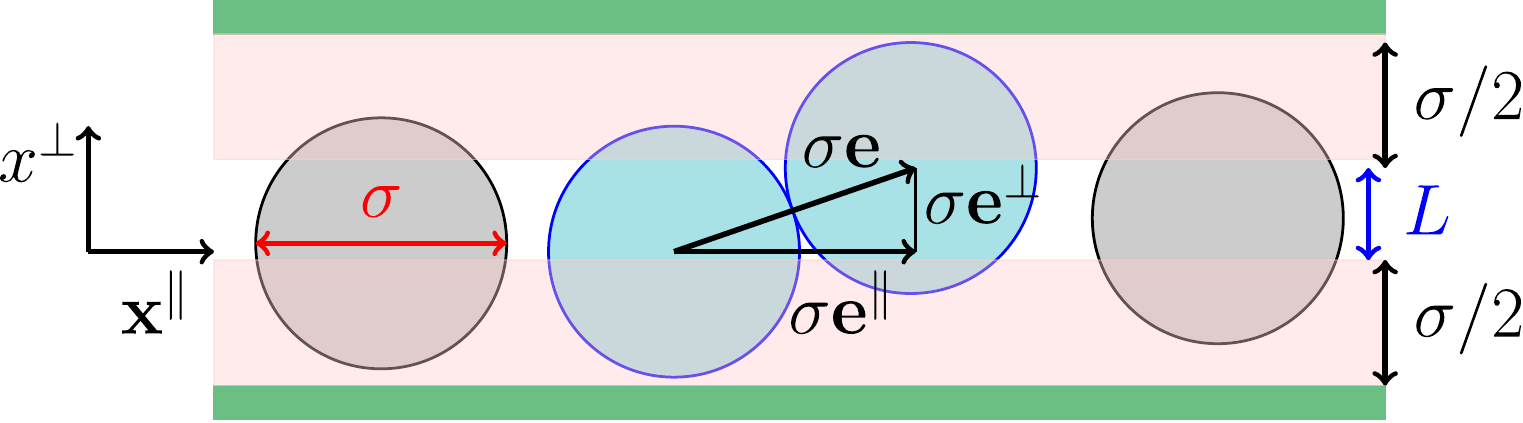}
\caption{Schematic cross section of hard spheres of diameter $\sigma$ confined to a slit of accessible width $L$. For two colliding spheres the velocity transfer is directed along the connecting vector $\sigma \vec{e}$ which lies almost parallel to the walls for small widths $L$. }
\label{fig:drawing}

\end{figure}

In this Letter, we now address the dynamical confinement problem  and demonstrate that as the fluid becomes more and more confined a singular time scale emerges separating 2D from 3D behavior. The dependence of this divergent time scale on the plate separation and the packing fraction will be worked out analytically including the prefactor and validated by simulations.

\paragraph{Simulation}

We investigate a fluid of  hard spheres of diameter $\sigma$ confined between two flat parallel hard walls 
with accessible slit width $L$, see Fig.~\ref{fig:drawing}. Trajectories are computed by event-driven molecular dynamics with initial velocities drawn from a Maxwell Boltzmann distribution with thermal velocity $v_{th} = \sqrt{k_B T/m}$, which also sets the natural time scale $t_0 = \sigma /v_{th}$.  
We focus on small wall separations $L/\sigma= 0.01, \ldots, 0.5 < 1 $ such that only a single monolayer fits between the plates and consider systems at  2D packing fractions 
$\varphi_{2D} = (\pi/4) \sigma^2 N/A = 0.1, \ldots, 0.8 $ for $N$ particles per area $A$, such that the highest densities are already beyond the freezing transition.

To unravel the divergent time scale we choose an observable that displays nontrivial dynamics exclusively due to the weak coupling of the 
transversal to the lateral degrees of freedom. Since in the decoupled ensemble the lateral degrees of freedom evolve just like in a confined ideal  gas, 
a  natural candidate is the 
 transversal kinetic energy $ \epsilon^{\perp}_s(t) = (m/2) [v_s^\perp(t)]^2$ of 
a tagged particle, i.e. any $s\in\{1,\ldots, N\}$, and   $v_s^\perp(t)$ is the fluctuating velocity perpendicular to the plates. We therefore monitor the time-correlation function 
\begin{equation}
 T_s^\perp(t) = \langle \delta \epsilon^\perp_s(t) \delta \epsilon^\perp_s(0) \rangle,
\end{equation}
of the fluctuations $\delta \epsilon^\perp_s(t) = \epsilon^\perp_s(t) - k_B T/2$. From the Maxwell-Boltzmann distribution one readily computes its initial value $T_s^\perp(0) = \langle |\delta \epsilon^\perp_s |^2 \rangle = (k_B T)^2/2$. 
 
\begin{figure}
\includegraphics*[width=\linewidth]{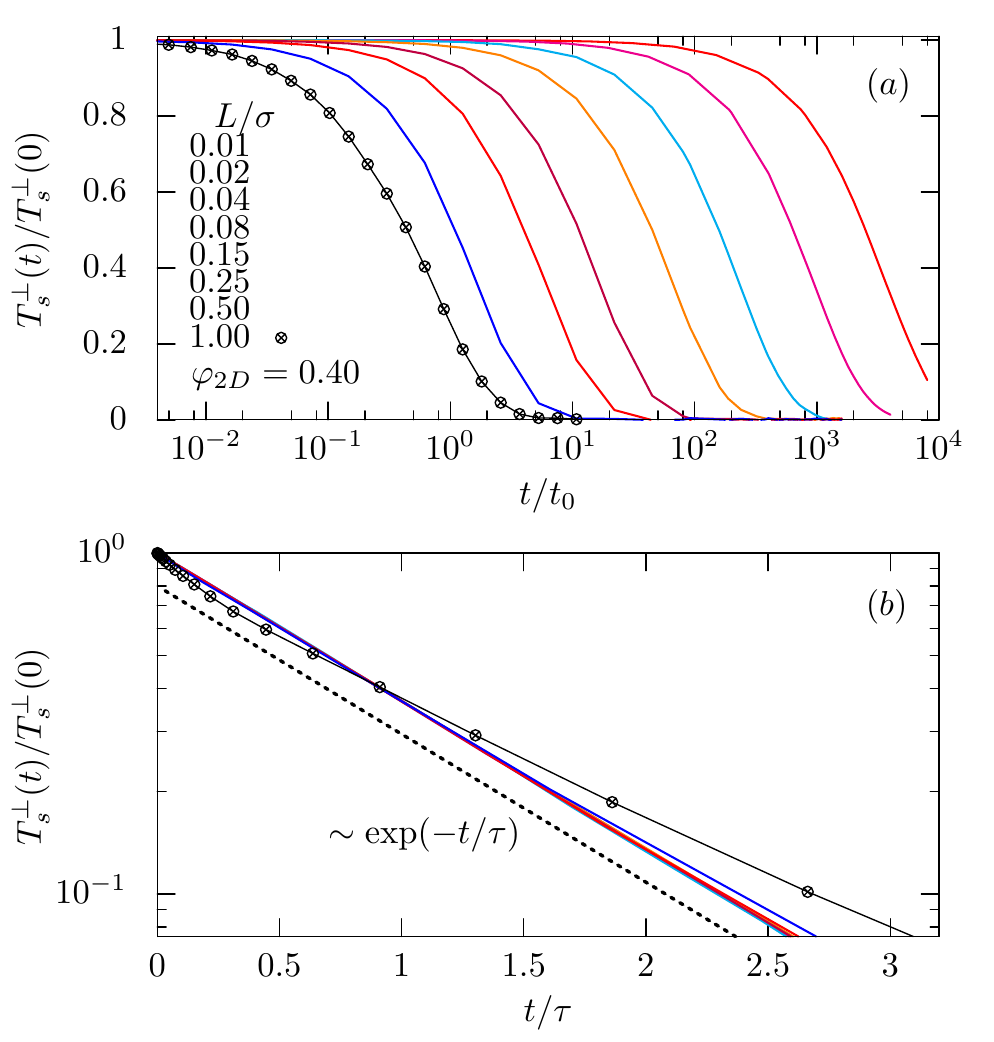}
\caption{(a) Decay of the time-correlation function  $T_{s}^\perp(t)$ of the transversal kinetic energy for $\varphi_{2D}=0.40$ and various
 wall separations $L$. Wall distance decreases from left to right.  (b) Same data in semi-log plot after rescaling with the measured relaxation time $\tau$. As a guide to the eye, the dashed line indicates a pure exponential decay $\sim \exp(-t/\tau)$.}
\label{fig:decay}
\end{figure}

The simulation results are displayed in Fig.~\ref{fig:decay}(a) for the moderate packing fraction $\varphi_{2D}=0.4$ (well below the two-dimensional freezing transition to a triangular phase $\phi_{\text{2D}}^\text{freezing} \approx 0.69$ \cite{Schmidt:1996,Fortini:2006,Franosch:2012})   for accessible plate separations covering two decades. One infers that the characteristic time scale increases by 4 orders of magnitude as the plate separation is decreased by a factor of 100, while the shape of the relaxation function becomes independent of the plate distance for small $L$. This suggests that data collapse can be achieved upon proper rescaling with the measured relaxation time.  Plotting the data on a semi-log plot 
then demonstrates that for small $L$ the data follow a pure exponential $\exp(-t/\tau)$ even at the smallest time scales, see Fig.~\ref{fig:decay}(b).  
Deviations become apparent only at the largest distance  considered, $L/\sigma=1$.

\begin{figure}
\includegraphics*[width=0.9\linewidth]{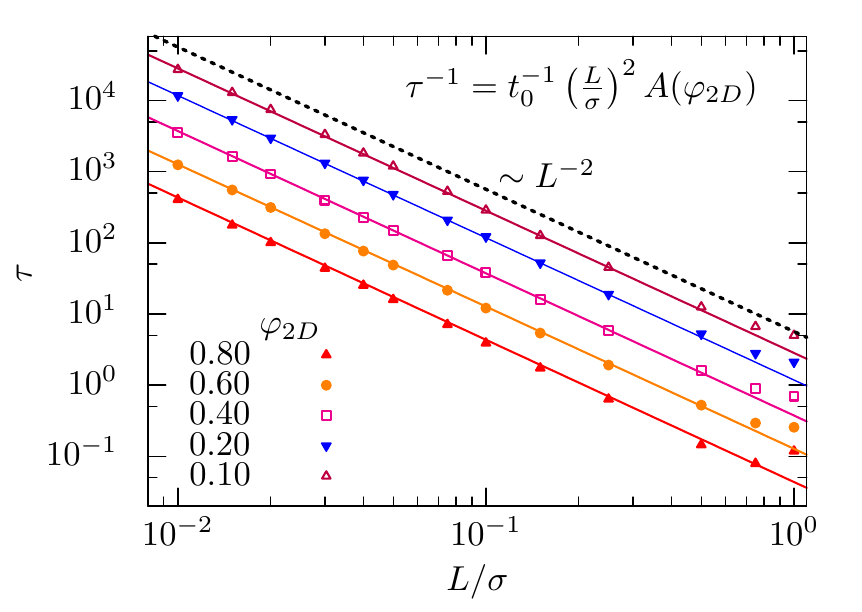}
\caption{Relaxation time $\tau$ for the strongly confined liquids as function of the wall separation $L$ for various packing fractions~$\varphi_{2D}$. Packing fraction increases from top to bottom. The dashed line is a power law $\sim L^{-2}$ and serves as a guide to the eye.}
\label{fig:relax}
\end{figure}

\begin{figure}
\includegraphics*[width=0.9\linewidth]{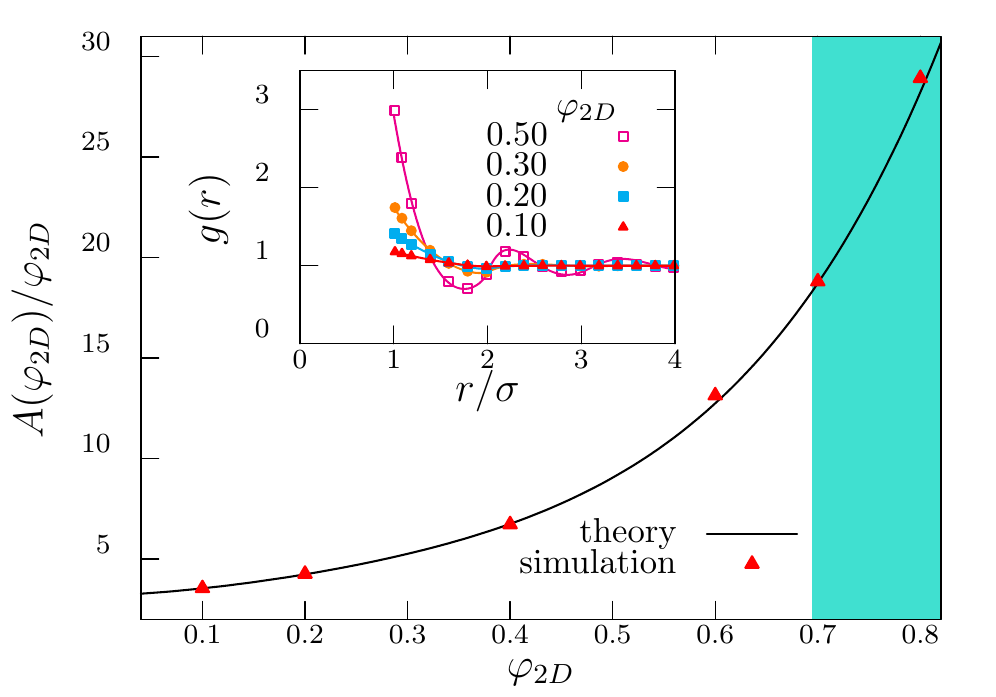}
\caption{ Prefactor $A(\varphi_{2D})$ for the scaling behavior of the relaxation rate $\tau(L,\varphi_{2D})^{-1} = t_0^{-1} (L/\sigma)^2 A(\varphi_{2D})$ as a function of $2D$ packing fraction $\varphi_{2D}$. The shaded area indicates the ordered phase. The inset shows the $2D$ radial distribution function at some representative packing fractions.  
The solid lines are from the Ornstein-Zernike relation using the Percus-Yevick closure, 
whereas the symbols are from simulations. }
\label{fig:rdf}
\end{figure}

The relaxation times $\tau = \tau(L,\varphi_{2D})$ extracted from the simulations, see Fig.~\ref{fig:relax}, approach a divergence $\sim L^{-2}$ for all packing fractions considered, the power law being an excellent description of the data already at wall separations $L/\sigma \lesssim 0.5$. 
Dimensional analysis suggests that in this regime the relaxation rate should read
\begin{equation}\label{eq:relaxation_rate}
 \tau^{-1} = t_0^{-1} \left( \frac{L}{\sigma} \right)^2  A(\varphi_{2D})  ,
\end{equation}
where the prefactor $A(\varphi_{2D})$ depends on the packing fraction only. The prefactors $A(\varphi_{2D})$ as measured from the simulation data, displayed in Fig.~\ref{fig:rdf}, increases smoothly with the packing fraction, in particular, it grows stronger than the packing fraction itself.

\paragraph{Theory}
The coordinates and velocities of the particles split naturally into lateral and transversal degrees of freedom, $\vec{x}_n = ( \vec{x}_n^\parallel, x_n^\perp)$, $\vec{v}_n = ( \vec{v}_n^\parallel, v_n^\perp)$, $n=1,\ldots,N$, the confinement restricts the transversal coordinates to $|x_n^\perp| \leq L/2$, see Fig.~\ref{fig:drawing}. The time evolution of an observable $A(t) = A( \{\vec{x}_n(t)\}, \{\vec{v}_n(t) \})$ is inherited from the trajectories and is  formally encoded in $ A(t) = \exp( i {\cal L}_{\pm} t )  A,\, t \gtrless 0$, with the pseudo-Liouville operator~\cite{Ernst:1969,Kinetic_theory_of_granular_gases:2010}  
\begin{equation}
 i {\cal L}_\pm =  \sum_{n} \vec{v}_n \cdot \frac{\partial }{\partial \vec{x}_n} + \sum_{n} \hat{W}_\pm(n) + \sum_{m< n} \hat{T}_\pm(m,n) .  
\end{equation}
Here the first term describes merely the free-streaming motion, while the operator $\hat{W}_\pm(n)$ in the second term accounts  for the collision of particle $n$ with the hard walls. Explicit expressions are readily derived  following the standard method~\cite{Kinetic_theory_of_granular_gases:2010} but won't be needed in the following. 
The interaction between the hard spheres is encoded in the binary collision operator 
\begin{align}\label{eq:collision_operator}
 \hat{T}_\pm(mn) = \sigma^2 \! \int \! \diff \vec{e} \, \Theta(\mp \vec{v}_{mn} \cdot \vec{e}) |\vec{v}_{mn}\cdot \vec{e}| \nonumber \\ \times \delta^{(3)}(\vec{x}_{mn}-\sigma\vec{e} ) \left[ \hat{b}^{\vec{e}}(mn) -1 \right] .
\end{align}
Here $\vec{e}$ is a unit vector and the integral extends over the unit sphere. 
The normal component of the relative velocity of the colliding pair $\vec{v}_{mn}\cdot \vec{e} = (\vec{v}_m-\vec{v}_n) \cdot \vec{e}$ (multiplied by the infinitesimal time $\diff t$) defines the collision cylinder. The Heaviside step function $\Theta(\cdot)$ selects approaching/distancing particles and the $\delta$-function ensures the contact condition for the collision and determines the unit vector $\vec{e}$. The operator $\hat{b}^{\vec{e}}(mn)$ acts only on the velocities $\vec{v}_m$ and $\vec{v}_n$ and replaces them by the velocities $\tilde{\vec{v}}_m, \tilde{\vec{v}}_n$ after the collision
\begin{align}\label{eq:collision_rule}
 b^{\vec{e}}(mn) \vec{v}_m &:= \tilde{\vec{v}}_m = \vec{v}_m - \vec{e} (\vec{e} \cdot \vec{v}_{mn} ) ,\nonumber \\
 b^{\vec{e}}(mn) \vec{v}_n &:= \tilde{\vec{v}}_n = \vec{v}_n + \vec{e} (\vec{e} \cdot \vec{v}_{mn} ) .
\end{align}

The coupling between the transversal and lateral degrees of freedom occurs only via the collisions $\hat{T}_\pm(mn)$. Yet, if the accessible slit width  is much smaller than the hard-sphere diameter, $L \ll \sigma$, the unit vector $\vec{e} = (\vec{e}^\parallel, e^\perp)$ for the contact condition satisfies $|e^\perp| \leq L/\sigma \ll 1, |\vec{e}^\parallel| \simeq 1$,   see Fig.~\ref{fig:drawing}, such that the momentum transfer is almost planar. This insight suggests that the collision operator may be replaced to leading order by its two-dimensional analogue $\hat{T}_\pm^\parallel(mn)$ where the contact condition 
$\delta^{(3)}(\vec{x}_{mn}-\sigma \vec{e}) \mapsto \delta^{(2)}(\vec{x}_{mn}^\parallel-\sigma \vec{e}^\parallel) \delta(\sigma e^\perp)$ 
 in Eq.~\eqref{eq:collision_operator} is replaced by a collision within the plane. 
Then the pseudo-Liouville operator naturally decomposes
\begin{equation}
 i {\cal L}_\pm = i{\cal L}^\perp_{\pm} + i {\cal L}^\parallel_{\pm}  + i {\cal L}^{\text{int}}_\pm ,
\end{equation}
where ${\cal L}_\pm^\perp$ accounts for the transversal degrees of freedom of a confined ideal gas, ${\cal L}_\pm^\parallel$ corresponds to the time evolution of a two-dimensional hard-disk fluid, while the residual interaction is encoded in
\begin{align}
i {\cal L}_\pm^{\text{int}} =& \sum_{m<n} \left[ \hat{T}_\pm(m,n) - \hat{T}_\pm^\parallel(m,n) \right] . 
\end{align}
The idea  is that $i {\cal L}_\pm^{\text{int}}$ induces only a weak coupling. Ignoring this contribution leads to a decoupled ensemble of interacting lateral degrees of freedom and an ideal gas of transversal degrees of freedom. This weak coupling therefore introduces a time scale up to which the coupling of the degrees of freedom is irrelevant. As the plate separation becomes smaller this time scale is expected to grow. 
 
The transversal kinetic energy  is conserved in the decoupled ensemble $\partial_t \delta \epsilon_s^\perp = i {\cal L}^{\text{int}}_\pm \delta \epsilon_s^\perp$  and the decay of its time correlation function $T^\perp_s(t)$ directly reflects the small coupling. Relying on the Zwanzig-Mori projection operator formalism~\cite{Hansen:Theory_of_Simple_Liquids,Goetze_Book_Complex_Dynamics} an exact equation of motion (e.o.m) can be readily derived~\cite{supplement_slowtime} for $t> 0$
\begin{equation}\label{eq:eom}
 \dot{T}_s^\perp(t) +  \tau^{-1} T_s^\perp(t) + \int_0^t  K^\perp_s(t-t') T_s^\perp(t') \diff t' = 0.
\end{equation}
Here, the second term describes an instantaneous relaxation, whereas the convolution integral accounts for the retarded friction due to correlated sequences of collisions. 
A short-time expansion of the e.o.m, Eq.~\eqref{eq:eom}, yields $T_s^\perp(t)/T_s^\perp(0) = 1- t/\tau + {\cal O}(t^2)$, and the convolution integral over the memory kernel $K^\perp_s(t)$ contributes only to order ${\cal O}(t^2)$. Direct expansion  of $\delta \epsilon_s^\perp(t) = \exp( i {\cal L}_\pm t) \delta \epsilon_s^\perp$ in powers of $t$ in $T_s^\perp(t)$ yields for the relaxation rate the microscopic expression
\begin{equation}\label{eq:relax}
 \tau^{-1} = \langle \delta \epsilon^\perp_s   (\mp i {\cal L}_\pm^{\text{int}} \delta \epsilon^\perp_s )  \rangle \frac{2}{(k_B T)^2}.
\end{equation}
For the memory kernel a microscopic expression also follows 
\begin{equation}
 K^\perp_s(t) = \langle {\cal Q} {\cal L}_+^{\text{int}}\delta \epsilon^\perp_s | e^{-i {\cal Q} {\cal L}_- {\cal Q} t} | {\cal Q} {\cal L}_-^{\text{int}} \delta \epsilon^\perp_s  \rangle \frac{2}{(k_B T)^2} ,
\end{equation}
where ${\cal Q}$ projects onto the subspace orthogonal to $\delta \epsilon_s^\perp$ and the bracket corresponds to Kubo's scalar product~\cite{Hansen:Theory_of_Simple_Liquids,Goetze_Book_Complex_Dynamics,supplement_slowtime}. 
Since the memory kernel contains at least two operators ${\cal L}_\pm^{\text{int}}$ which we identified as small perturbation, the retarded convolution term in Eq.~\eqref{eq:eom} becomes negligible with respect to the instantaneous relaxation. Then the e.o.m. [Eq.~\eqref{eq:eom}] simplifies to an exponential relaxation as leading behavior consistent with our simulation results. 

The relaxation rate, Eq.~\eqref{eq:relax}, involves an equilibrium average of essentially the collision operator. Its direct evaluation becomes feasible  for strong confinement relying  on the decoupling property of the ensemble into lateral and transversal degrees of freedom, as well as the usual decoupling of the structural and kinetic degrees of freedom.  The second crucial ingredient is  that the unit vector $\vec{e} = (\vec{e}^\parallel, e^\perp)$ for the contact condition becomes more and more confined to the planar direction, $|e^\perp| \leq L/\sigma \ll 1$. By the collision rule, Eq.~\eqref{eq:collision_rule}, $\tilde{v}_s^\perp 
= v_s^\perp + e^\perp (\vec{e} \cdot \vec{v}_{ms})$ the transverse velocity   remains almost unchanged after a collision. After performing the structural and kinetic averages we obtain as leading contribution (see Supplemental Material~\cite{[{See Supplemental Material which includes Ref.~\cite{Bollinger:2016,Bollinger:2015,Weeks:1971,Plimpton:1995,Tonks:1936}, for details on the Zwanzig-Mori equations and smooth potentials}]supplement_slowtime})
\begin{equation}\label{eq:rate_analytic}
 \tau^{-1} = \frac{16 \varphi_{2D} }{3 \sqrt{\pi} } \left(\frac{L}{\sigma}\right)^2    g(\sigma) t_0^{-1},  
\end{equation}
where $g(\sigma)$ the radial distribution function of the two-dimensional hard-disk fluid at contact. The factor $\varphi_{2D} g(\sigma)$ accounts 
for the probability of a scattering event similar to Enskog's theory for bulk 
hard-sphere fluid~\cite{Hansen:Theory_of_Simple_Liquids}. In contrast to the 3D case here it arises as an 
exact result valid at any packing fraction $\varphi_{2D}$.  The decoupling of the lateral and transversal degrees of freedom is encoded in the factor $(L/\sigma)^2$ reflecting the small momentum transfer in quasi-planar collisions.  

The radial distribution function $g(r)$ can be evaluated  within  integral equations theory~\cite{Hansen:Theory_of_Simple_Liquids}. Here we rely on a numerical solution of the 2D Percus-Yevick closure relation~\cite{Adda-Bedia:2008} that compares quantitatively to our measured $g(r)$ in the simulation, see Fig.~\ref{fig:rdf}. Using the contact value $g(\sigma)$ the prefactor $A(\varphi_{2D})$ for the relaxation rate in Eq.~\eqref{eq:relaxation_rate} can be compared to the analytic result, Eq.~\eqref{eq:rate_analytic}. The comparison in Fig.~\ref{fig:rdf} for low to moderate packing fractions $\varphi_{2D}$ corroborates that the theoretical prediction is in fact an exact result.

\paragraph{Conclusions} We have demonstrated the emergence of a divergent time scale for the coupling of lateral to transverse degrees of freedom in a strongly confined fluid. The main insight has been that in collisions the momentum transfer becomes more and more planar as the wall separation is reduced. The dependence of the divergent time scale on the plate separation and packing fraction of the fluid has been worked out analytically including the prefactor by evaluating the dominant contribution of the collision operator and compared to our simulations. Remarkably, the theory is not limited to  fluids but also applies to the ordered phase. We emphasize that the reference system is strongly interacting and our calculation is one of the rare cases where analytic results can be elaborated. 

The mechanism unraveled for the emergence of a slow time scale and the scaling with the transverse dimension should also hold for other geometries such as liquids in narrow cylindrical tubes 
or quasi-1D confinement (see Supplemental Material~\cite{supplement_slowtime}).

The hard-core interaction is of course an idealization of a short-ranged potential, but we anticipate our results to remain valid for the case of smooth potentials. More precisely, for hard spheres the collisions are instantaneous whereas for smooth potentials the duration of a collision introduces a new time scale into the problem. As long as the Knudsen time scale $L/v_{th}$, i.e. the typical time for a particle to traverse the slit, is still much larger than the duration of a collision the mechanisms for small transverse momentum transfer should be identical (see Supplementary Material for simulation results on smooth potentials~\cite{supplement_slowtime}). Similarly, a smooth particle-wall interaction should not modify our findings, provided its  range is much smaller than the slit width, and the transverse  energy includes the wall potential in addition to the transverse kinetic energy.  

It is also of interest to consider the opposite case where the duration of a collision is much longer than the time to traverse the slit. 
Then the use of  a collision operator is no longer justified; rather, the collision events can be averaged over the fast transverse oscillations. Analytic progress in this direction has been made very recently~\cite{Schilling:2016} and it turns out that the predicted relaxation time for this case scales with a different power in the wall separation. Furthermore, the relaxation of the kinetic energy becomes exponential at times much longer than the Knudsen time, while for hard spheres it is exponential for all times. 

The diverging relaxation time separates the decoupled two-dimensional dynamics from the coupled one in strong confinement. This should have 
drastic implications for systems in the vicinity of the glass transition such that the divergent structural relaxation time competes with
the relaxation time of the coupling. In fact the mode-coupling theory for confinement~\cite{Lang:2014a} suggests that the limits $t\to \infty$ and $L\to 0$ do not commute and different glassy dynamics on different time scales is expected.

The decoupling property of the transverse and lateral degrees of freedom in the equilibrium ensemble implies a divergent time scale for their dynamic coupling. The precise form of the divergence should depend on the microdynamics and should be different for the case of Brownian dynamics, which can be realized experimentally for colloids confined between glass plates.  Yet, to measure the divergent time scale in this case an observable needs to be chosen that does not relax quickly to equilibrium even without the close-to-planar collisions. 
An example could be the in-plane self-intermediate scattering function $F^s(q,t)$ at small wave numbers $q$, which probes the planar dynamics at large lateral length scales $2\pi/q$. Upon decreasing the wavenumber the relaxation time slows down as $\sim q^{-2}$ by diffusion and the crossover  from  purely 2D motion to the 3D confined  coupled dynamics should be visible. A second, more challenging candidate  for such an  observable 
is the generalized intermediate scattering functions for fluids close to the glass transition.

\begin{acknowledgments}
We gratefully acknowledge many discussions with Rolf Schilling on our simulation results and on the emergence of the diverging time scale.  
This work has been supported by the Deutsche Forschungsgemeinschaft DFG via the  Research Unit FOR1394 ``Nonlinear Response to
Probe Vitrification''.  
\end{acknowledgments}

%

\onecolumngrid
\clearpage

\section{Supplemental Material}
\section*{1. Zwanzig-Mori equations of motion and Relaxation time}

Consider the fluctuations of the transverse kinetic energy of a a tagged particle
\begin{equation}
 \delta \epsilon^\perp_s := \frac{m}{2} (v_s^\perp)^2 - \frac{1}{2} k_B T,
\end{equation}
 where the label $s$ is any of the $1,\ldots, N$ and $v_s^\perp$ the velocity component perpendicular to the walls, $m$ the mass of the particle and $k_B T$ the thermal energy. By equipartion $\langle \delta \epsilon^\perp_s \rangle = 0$. In the limit of small wall separation the ensemble decouples and $\delta \epsilon^\perp_s$ becomes a conserved quantity. Here we derive an exact equation of motion for the time-correlation function
of  $\epsilon^\perp_s(t) = e^{i {\cal L}_\pm t} \epsilon^\perp_s$  
\begin{equation}
 T^\perp_s(t) := \langle \delta \epsilon^\perp_s(t) \delta \epsilon^\perp_s \rangle = \langle e^{i {\cal L}_\pm t} \delta \epsilon^\perp_s | \delta \epsilon^\perp_s \rangle = \langle \delta \epsilon^\perp_s | e^{-i {\cal L}_{\mp} t } \delta \epsilon^\perp_s \rangle, \qquad t \gtrless 0.
\end{equation}
Here $\langle A | B \rangle \equiv \langle A^* B \rangle$ abbreviates the Kubo scalar product~\cite{Goetze_Book_Complex_Dynamics} and the adjoint of the pseudo-Liouvillian with respect to this scalar product fulfills $({\cal L}_\pm)^\dagger = {\cal L}_\pm$. 
In the following we restrict the discussion to  non-negative times $t\geq 0$ only, negative times follow trivially. The initial value is readily calculated since the momenta are drawn from a Maxwell-Boltzmann  distribution
\begin{equation}\label{eq:normalization}
 T^\perp_s(t=0) = \langle | \delta \epsilon_s^\perp |^2 \rangle = \frac{1}{2} (k_B T)^2 .
\end{equation}

We rely on the operator identity for the backwards-time evolution operator ${\cal R}(t) = \exp(-i {\cal L}_\mp t), t \gtrless 0$
\begin{align}\label{eq:operator_identity}
 \partial_t {\cal P} {\cal R}(t) {\cal P} + i {\cal P} {\cal L}_\mp {\cal P} {\cal R}(t) {\cal P} + \int_0^t \diff t' \, {\cal P} {\cal L}_\mp {\cal Q} {\cal R}_{\cal Q}(t-t') {\cal Q} {\cal L}_\mp {\cal P} {\cal R}(t) {\cal P}=0 ,
\end{align}
valid for any orthogonal projector ${\cal P}$, see e.g. \cite{Lang:2012}. Here ${\cal Q} = \mathbf{1} - {\cal P}$ denotes the projection onto the orthogonal complement, and ${\cal R}_{\cal Q}(t) = \exp(-i {\cal Q}{\cal L}_\mp {\cal Q} t)$ is referred to as the reducd backwards-time evolution operator. 

For the projector we choose
\begin{equation}
 {\cal P} = | \delta \epsilon^\perp_s \rangle \frac{2}{(k_B T)^2}  \langle \delta \epsilon^\perp_s | ,
\end{equation}
which by the normalization Eq.~\eqref{eq:normalization} indeed fulfills ${\cal P} = {\cal P}^2 = {\cal P}^\dagger$. From the operator identity, Eq.~\eqref{eq:operator_identity} one finds the exact Zwanzig-Mori equation of motion
\begin{equation}
 \dot{T}^\perp_s(t) \pm  \tau^{-1} T^\perp_s(t) + \int_0^t  K^\perp_s(t-t') T^\perp_s(t') \diff t' = 0, \qquad t\gtrless 0
\end{equation}
An explicit expression for the characteristic relaxation rate $\tau^{-1}$ follows
\begin{equation}\label{eq:relaxation_rate_explicit}
 \tau^{-1} = \langle \delta \epsilon^\perp_s |\pm  i {\cal L}_\mp^{\text{int}} \delta \epsilon^\perp_s \rangle \frac{2}{(k_B T)^2} , 
\end{equation}
and similarly for  the memory kernel
\begin{equation}
 K^\perp_s(t) = \langle {\cal Q} {\cal L}_\pm^{\text{int}} \delta \epsilon^\perp_s | e^{-i {\cal Q} {\cal L}_\mp {\cal Q} t} | {\cal Q} {\cal L}_\mp^{\text{int}} \delta \epsilon^\perp_s  \rangle \frac{2}{(k_B T)^2} , \qquad t \gtrless 0,
\end{equation}
where we observed ${\cal L}_\pm \delta \epsilon_s^\perp = {\cal L}_\pm^{\text{int}} \epsilon_s^\perp$. 

Here we focus on the evaluation of the relaxation rate $\tau^{-1}$ of the transeverse kinetic energy, Eq.~\eqref{eq:relaxation_rate_explicit}. Since the transverse dynamics conserves the transverse kinetic energy and $\delta \epsilon_s^\perp$ does not involve lateral degrees of freedom, only collisions contribute to the relevant matrix element
\begin{align}
& \langle \delta \epsilon_s^\perp | i{\cal L}_-^{\text{int}} \delta \epsilon_s^\perp \rangle = 
\langle \delta \epsilon_s^\perp | \sum_{ m=1,m\neq s}^N  \delta \hat{T}_-(m,s) \delta \epsilon_s^\perp \rangle  \nonumber \\
=&  \sum_{\stackrel{ m=1}{m\neq s}}^N  \left\langle \sigma^2 \int \diff \vec{e}\,  \Theta(\vec{v}_{ms} \cdot \vec{e} ) |\vec{v}_{ms} \cdot \vec{e} | \left[ \delta^{(3)}(\vec{x}_{ms} - \sigma \vec{e}) -\delta^{(2)}(\vec{x}_{ms}^\parallel - \sigma \vec{e}^\parallel) \delta(\sigma e^\perp)
 \right] \left[\frac{m}{2} (v_s^\perp)^2 - \frac{1}{2} k_B T \right]  \left[\frac{m}{2} (\tilde{v}_s^\perp)^2 
- \frac{m}{2}(v_s^\perp)^2 \right]  \right\rangle ,
\end{align}
with the post-collision velocity $\tilde{\vec{v}}_s =( \tilde{\vec{v}}_s^\parallel, \tilde{v}_s^\perp) = \vec{v}_s + \vec{e} (\vec{e} \cdot \vec{v}_{ms} ) $ of the tagged particle. Since for purely planar collisions the transverse velocity does not change the second $\delta$-functions does not contribute. 

For small wall separations, Fig.~1 of the main text reveals that  the unit vector $\vec{e} = (\vec{e}^\parallel,e^\perp)$ is confined to a small region around the equator $|e^\perp| \leq L/ \sigma \ll 1, |\vec{e}^\parallel| \simeq 1$. Furthermore the thermal average over the structural degrees of freedom can be performed observing that in the reference ensemble the degrees of freedom decouple into a two-dimensional interacting fluid for the lateral degrees of freedom and an ideal gas for the transverse ones~\cite{Franosch:2012,Lang:2014}. 
\begin{align}
 \langle \delta \epsilon_s^\perp | i{\cal L}_-^{\text{int}} \delta \epsilon_s^\perp \rangle 
=&   n_0 \sigma^2 g(\sigma) \int \diff \vec{e}\,   \Big\langle  \Theta(\vec{v}_{ms} \cdot \vec{e} ) |\vec{v}_{ms} \cdot \vec{e} |  \left[\frac{m}{2} (v_s^\perp)^2 - \frac{1}{2} k_B T \right]  \left[\frac{m}{2} (\tilde{v}_s^\perp)^2 - \frac{m}{2}(v_s^\perp)^2 \right] \Big\rangle 
 \nonumber \\
& \times \int \frac{\diff z_m}{L}  \int \frac{\diff z_s}{L}   \delta(z_m -z_s - \sigma e^\perp)   .
\end{align}
Here $n_0=N/A$ is the two-dimensional particle density and $g(\sigma)$ the radial distribution function of the two-dimensional reference fluid at contact. 
The integral in the preceeding equations over the transverse degrees of freedom extends from $-L/2$ to $L/2$. By a change of variables $z_{ms} =z_m-z_s$
\begin{align}
 \int_{-L/2}^{L/2} \diff z_m \int_{-L/2}^{L/2} \diff z_s   \delta(z_m -z_s - \sigma e^\perp)  =& \int_{-L}^{L} \diff z_{ms} \delta(z_{ms} - \sigma e^\perp)  \int_{z_s \in [-L/2,L/2], z_s + z_{ms} \in [-L/2,L/2]} \diff z_s \nonumber \\
=& \int_{-L}^{L} \diff z_{ms}  \delta(z_{ms}- \sigma e^\perp)  (L- | z_{ms}|) \nonumber \\
=& L  - \sigma |e^\perp|,
\end{align}
and we understand that from now on $|e^\perp| \leq L/\sigma$.  
The integral over the orientation of the contact point thus collapses to a small margin at the equator:
\begin{equation}
 \int \diff \vec{e}\,  \Theta( \vec{v}_{ms}\cdot \vec{e} ) \ldots \mapsto \int_{-L/\sigma}^{L/\sigma} \diff e^\perp \int \diff \vec{e}^\parallel\,  \Theta(\vec{v}_{ms}^\parallel \cdot \vec{e}^\parallel ) \ldots \quad ,
\end{equation}
where $e^\perp$ is  the latitude on the unit sphere and the integral $\diff \vec{e}^\parallel$ extends over a unit circle. Then the matrix element further simplifies to 
\begin{align}\label{eq:matrix_element_kinetic}
& \langle \delta \epsilon_s^\perp | i{\cal L}_-^{\text{int}} \delta \epsilon_s^\perp \rangle  = \nonumber \\ 
&  = n_0 \sigma  g(\sigma)   \int_{-L/\sigma}^{L/\sigma} \left( 1- \frac{\sigma}{L} |e^\perp| \right) \frac{\sigma}{L} \diff e^\perp 
\int\!\diff \vec{e}^\parallel 
\, \Big\langle \Theta(\vec{v}_{ms}^\parallel \cdot \vec{e}^\parallel ) (\vec{v}_{ms} \cdot \vec{e} )   \left[\frac{m}{2} (v_s^\perp)^2 - \frac{1}{2} k_B T \right]  \left[\frac{m}{2} (\tilde{v}_s^\perp)^2 - \frac{m}{2}(v_s^\perp)^2 \right]  \Big\rangle .
\end{align}
After integrating over the latitude $e^\perp$ only even powers in the thermal average $\langle \ldots \rangle$ contribute. Furthermore since the contact point are close to the equator we need only the lowest even contribution in $e^\perp$. 
Using  the collision rule 
\begin{equation}
 \tilde{v}_s^\perp = v_s^\perp + e^\perp (\vec{v}_{ms} \cdot \vec{e}) = v_s^\perp + e^\perp (\vec{v}_{ms}^\parallel \cdot \vec{e}^\parallel) + e^\perp v_{ms}^\perp e^\perp,
\end{equation}
the thermal average in the bracket of Eq.~\eqref{eq:matrix_element_kinetic} reduces to (keeping in each step only the even leading terms in $e^\perp$)
\begin{align}
\langle \ldots \rangle =& \Big\langle  \Theta(\vec{v}_{ms}^\parallel \cdot \vec{e}^\parallel ) 
\,     \left[\frac{m}{2} (v_s^\perp)^2 - \frac{1}{2} k_B T \right]  
\frac{m}{2} \left[  2 v_s^\perp e^\perp (\vec{v}_{ms}\cdot \vec{e})^2 + (e^\perp)^2 (\vec{v}_{ms}\cdot \vec{e} )^3  \right]  \Big\rangle \nonumber \\
= & \Big\langle \Theta(\vec{v}_{ms}^\parallel \cdot \vec{e}^\parallel )\left[\frac{m}{2} (v_s^\perp)^2 - \frac{1}{2} k_B T \right]  
\frac{m}{2} \left[  4 (v_s^\perp e^\perp)^2 (\vec{v}_{ms}^\parallel \cdot \vec{e}^\parallel) + (e^\perp)^2 (\vec{v}_{ms}^\parallel \cdot \vec{e}^\parallel)^3  \right]  \Big\rangle ,
\end{align}
since $v_s^\perp$ is a gaussian variable the second term in the second bracket also does not contribute, and the remaining average factorizes into gaussian integrals 
\begin{align}
\langle \ldots \rangle = &  4 (e^\perp)^2 \langle \Theta(\vec{v}_{ms}^\parallel \cdot \vec{e}^\parallel ) \vec{v}_{ms}^\parallel \cdot \vec{e}^\parallel \rangle\Big\langle \left[\frac{m}{2} (v_s^\perp)^2 - \frac{1}{2} k_B T \right]  
\frac{m}{2}  (v_s^\perp)^2   \Big\rangle \nonumber \\
=& (e^\perp)^2 \frac{2 v_{th}}{\sqrt{\pi}} (k_B T)^2     .
\end{align}
Here we used the elementary result $\langle (\vec{v}_{ms}^\parallel \cdot \vec{e}^\parallel) \Theta(\vec{v}_{ms}^\parallel \cdot \vec{e}^\parallel  \rangle = v_{th} /\sqrt{\pi} $ for  the average relative velocity along the  approaching direction $\vec{e}^\parallel$. 
 With
\begin{equation}
 \int_{-L/\sigma}^{L/\sigma} \frac{\sigma}{L} \diff e^\perp \, (e^\perp)^2  \left( 1- \frac{\sigma}{L} |e^\perp| \right)    = \frac{L^2 }{6 \sigma^2} ,
\end{equation}
the remaining integrals in Eq.~\eqref{eq:matrix_element_kinetic} now yield
\begin{align}
 \langle \delta \epsilon_s^\perp | i{\cal L}_-^\text{int} \delta \epsilon_s^\perp \rangle 
=&   \frac{2 n_0}{3 \sigma} L^2  g(\sigma) (k_B T)^2 v_{th} \sqrt{\pi} .
\end{align}
For the relaxation rate in Eq.~\eqref{eq:relaxation_rate_explicit} this implies the principal result of this work
\begin{equation}\label{eq:final_result}
\boxed{
 \tau^{-1} = \frac{16 \varphi_\text{2D}}{3 \sqrt{\pi} } \left(\frac{L}{\sigma}\right)^2    g(\sigma) t_0^{-1} . 
}
\end{equation}
where we reintroduced the two-dimensional packing fraction $\varphi_\text{2D} = n_0 \pi \sigma^2/4$ and the natural time scale $t_0 = \sigma/v_{th}$


\section*{2. Smooth potentials}

\begin{figure}[h]
\includegraphics*[width=0.45\linewidth]{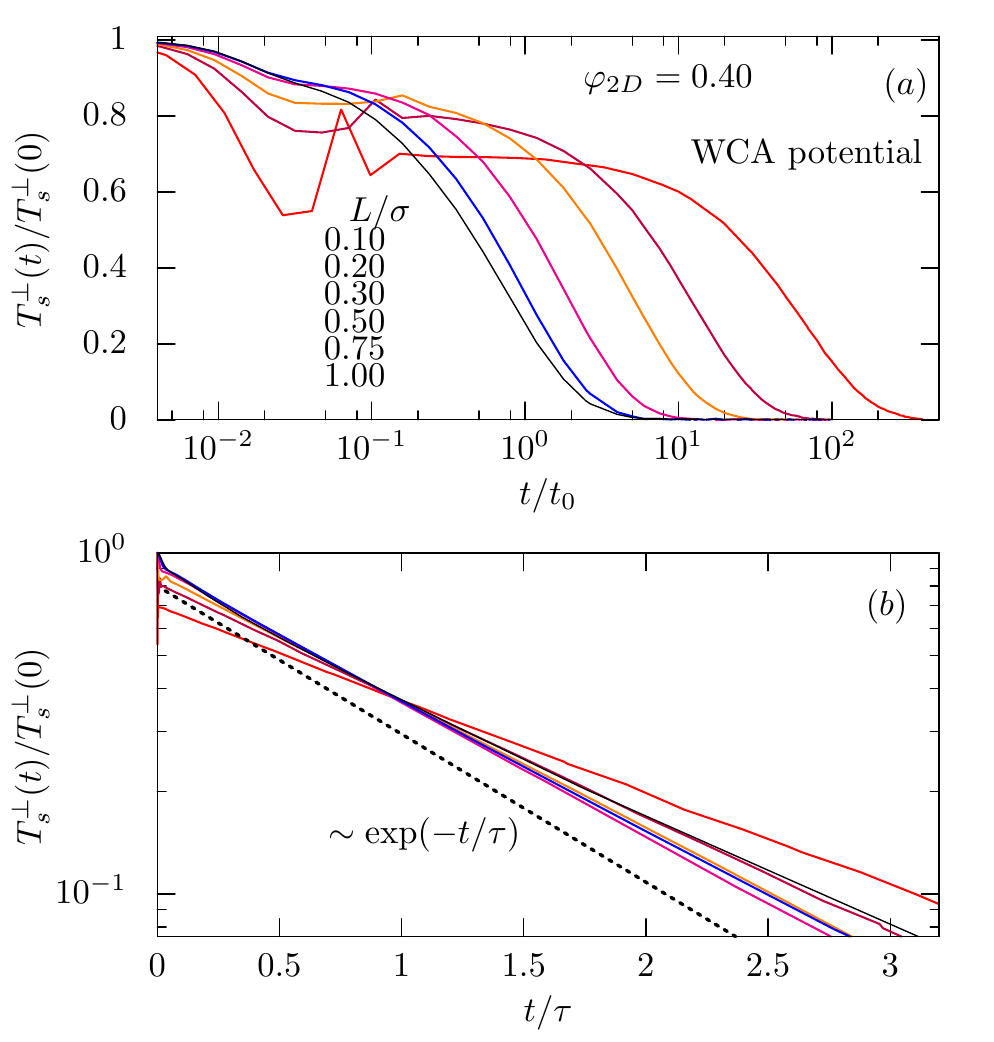} 
\caption{Decay of the time-correlation function $T_{s}^\perp(t)$ of the transversal kinetic energy of a liquid with smooth interactions for packing fraction  $\varphi_{2D}=0.40$ and various wall separations $L$. Wall distance decreases from left to right. (b) Same data in semi-log plot after rescaling with the measured relaxation time $\tau$.}
\label{fig:softExponential}
\end{figure}

We further simulate particles using a steep Weeks-Chandler-Andersen (WCA) potential as used in Ref.~\cite{Bollinger:2016,Bollinger:2015}, which captures many properties of dense atomistic liquids~\cite{Weeks:1971}. The particle-particle interactions are defined as $u_{pp}(r)=4 \epsilon[(\sigma/r)^{48} -(\sigma/r)^{24}] +\epsilon$ for $r \le 2^{1/24}\sigma$ and $u_{pp}(r)=0$ for $r \ge 2^{1/24}\sigma$, where $r$ and $\sigma$ are the inter-particle separation and particle diameter, respectively. For convenience, we use the characteristic energy scale $\epsilon=k_{B}T$, which sets the time scale $t_0=\sqrt{m\sigma^2/\epsilon}$. This liquid is confined to a slit  between two parallel and  flat walls placed at $\pm (\sigma +L/2)$ by using a steep WCA interactions, such that the accessible slit width is still given by $L$. Particles located inside the slit interact with  both walls, we employ the wall-particle interaction  $u_{wp}(z)=4 \epsilon[ (\sigma/z)^{48} -(\sigma/z)^{24}] +\epsilon$ for $z \le 2^{1/24}\sigma$ and $u_{wp}(z)=0$ for $z \ge 2^{1/24}\sigma$, where $z$ is the distance to one of the walls. We have performed  NVE molecular dynamics simulations using LAMMPS~\cite{Plimpton:1995}, and  integrate Newton's equations of motion with a time step of $ \delta t=10^{-5}t_0$. We equilibriate the systems for sufficintly long time such that 
on average each particle moves more than $10\sigma$.

\subsection{quasi-2D}

In order to quantify the divergent time scale, we calculate the time-correlation function 
$T_s^\perp(t) = \langle \delta \epsilon^\perp_s(t) \delta \epsilon^\perp_s(0) \rangle$ as shown in Fig.~\ref{fig:softExponential}. Since the potential is very steep it does not contribute significantly to the total transverse energy. One observes that the characteristic time scale increases by $2$ orders of magnitude as the plate separation is decreased by a factor of $10$. In contrast to hard sphere systems, these time-correlation functions are not purely exponential at very short time scales, which reflect that the collisions are no longer instantaneous. Yet, plotting the data on a semi-logarithmic plot reveals that the relaxation function becomes exponential at times much larger than the collision time.

\begin{figure}[h]
\includegraphics*[width=0.4\linewidth]{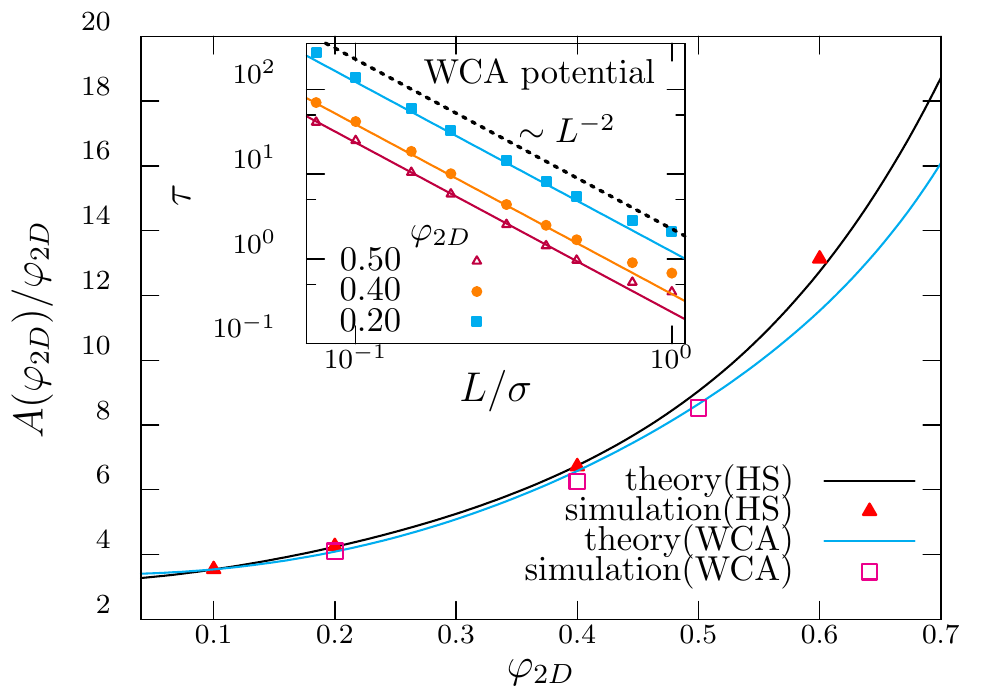}  
\caption{ Prefactor $A(\varphi_{\text{2D}})$ for  both the hard sphere system and the WCA liquid  extracted from  the scaling behavior of the relaxation time $\tau(L,\varphi_{2D})^{-1} = t_0^{-1} (L/\sigma)^2 A(\varphi_{2D})$ as a function of $2D$ packing fraction $\varphi_{2D}$. The inset shows 
the relaxation time $\tau$ for the strongly confined WCA  liquids as a function of the wall separation $L$ for various packing fractions~$\varphi_{2D}$. Packing fraction increases from top to bottom. The dashed line is a power law $\sim L^{-2}$ and serves as a guide to the eye.}
\label{fig:softtime_divergence}
\end{figure}

The relaxation time $\tau=\tau(L, \varphi_{\text{2D}})$ extracted from the WCA simulations, see the inset of Fig.~\ref{fig:softtime_divergence}, 
approaches the scaling law $\sim L^{-2}$ for all packing fractions considered. 
For comparison with Eq.~\eqref{eq:final_result} we use the maximum of the pair-distribution function $g(r)$ to replace  the contact value $g(\sigma)$ 
in the case of hard spheres. The prefactor of the scaling law is shown  Fig.~\ref{fig:softtime_divergence} for hard spheres and the WCA liquid for various densities and is described nicely by the theory.  Last, we have also tuned the particle-particle 
interactions of the WCA potential by changing the pair of  exponents from  ($48-24$) to  ($24-12$), but still observe the same divergence of the time scale $~\sim L^{-2}$.

\subsection{quasi-1D}

\begin{figure}[h]
\includegraphics*[width=0.4\linewidth]{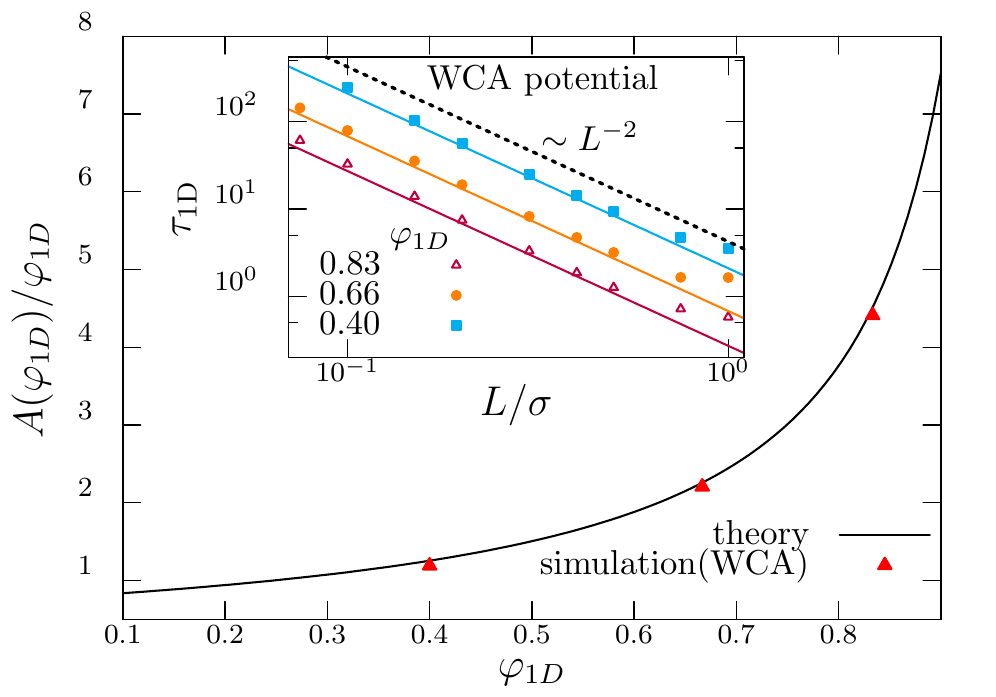}  
\caption{ Prefactor $A(\varphi_{\text{1D}})$ for the WCA liquid  extracted from  the scaling behavior of the relaxation time $\tau_{\text{1D}}^{-1}(L,\varphi_{1D})= t_0^{-1} (L/\sigma)^2 A(\varphi_{1D})$ as a function of $1D$ packing fraction $\varphi_{1D}$. The inset shows 
the relaxation time $\tau_{\text{1D}}$ for the strongly confined WCA  liquids as a function of the wall separation $L$ for various packing fractions~$\varphi_{1D}$. Packing fraction increases from top to bottom. The dashed line is a power law $\sim L^{-2}$ and serves as a guide to the eye.}
\label{fig:2Dchannels}
\end{figure}

We further simulate a two-dimensional WCA liquid in a narrow channel, and obtain the identical scaling exponent for different one-dimensional packing fractions $\varphi_{\text{1D}} = (N/L_c) \sigma$ for $N$ particles in a channel of lateral length $L_c >  N \sigma $ and accessible transverse slit width $L \ll \sigma$.  
The calculation for 
the relaxation rate can be adapted without conceptual changes also for narrow two-dimensional channels and yields
\begin{equation}\label{eq:final_result}
\boxed{
 \tau_{1D}^{-1} = \frac{4}{3 \sqrt{\pi}} \left(\frac{L}{\sigma}\right)^2  t_0^{-1} \frac{\varphi_\text{1D}}{1-\varphi_\text{1D}}. 
}
\end{equation}
Here we used the fact that the contact value  for a one-dimensional gas of hard particles is known exactly  $g(\sigma) = 1/(1-\varphi_{\text{1D}})$ 
from the Tonks gas~\cite{Tonks:1936}. 
The simulation results are compared to the theoretical prediction for the divergence of the coupling time scale  in the inset in  Fig.~\ref{fig:2Dchannels}.  The prefactor calculated from the theory is shown also in Figure~\ref{fig:2Dchannels} and corroborates nicely the theoretical expectation.

\end{document}